\documentclass[prb,floatfix,twocolumn,showpacs,amsmath,amssymb]{revtex4}
\usepackage{graphicx,color}
\usepackage{dcolumn}
\usepackage{bm}
\begin{document}
  
\title{Ground State of a Heisenberg Chain with Next-Nearest-Neighbor Bond Alternation}
\author{Luca Capriotti,$^{1}$ Federico Becca,$^{2,3}$  Sandro Sorella,$^{3}$
and Alberto Parola,$^{4}$}
\affiliation{
${^1}$ Kavli Institute for Theoretical Physics and Department of Physics, 
University of California, Santa Barbara CA 93106-4030 \\
${^2}$ Institut de Physique Th\'eorique, Universit\'e de Lausanne, CH-1015 Lausanne, Switzerland\\
${^3}$ INFM-DEMOCRITOS, National Simulation Centre, and SISSA, I-34014 Trieste, Italy\\ 
${^4}$ INFM and Dipartimento di Scienze, Universit\`a dell'Insubria, I-22100 Como, Italy }
    
\date{\today}
\begin{abstract}
We investigate the ground-state properties of the spin-half $J_1{-}J_2$ 
Heisenberg chain with a next-nearest-neighbor spin-Peierls dimerization using
conformal field theory and Lanczos exact diagonalizations.
In agreement with the results of a recent bosonization analysis by
Sarkar and Sen, we find that for small frustration ($J_2/J_1$) the system 
is in a Luttinger spin-fluid phase, with gapless excitations, 
and a finite spin-wave velocity. In the regime of strong frustration the 
ground state is spontaneously dimerized and the bond alternation reduces 
the triplet gap, leading to a slight enhancement of the critical point
separating the Luttinger phase from the gapped one. 
An accurate determination of the phase boundary is obtained numerically from 
the study of the excitation spectrum.
\end{abstract}

\maketitle

Asymmetric spin-half ladders have recently attracted 
much theoretical interest~\cite{voit,cinesini,sen,saw,naka} due to possible 
experimental realizations in delafossite cuprates such as ${\rm YCuO_{2.5}}$
(Ref.~\onlinecite{dela}) and the unusual physical effects that the asymmetry 
in the leg exchanges could introduce both in the ground-state correlations 
and in the properties of the excitation spectrum.
The simplest and most transparent Hamiltonian representation 
of these systems can be given in terms of frustrated ($J_1{-}J_2$) 
spin-half Heisenberg chains with a spin-Peierls dimerization
in the next-nearest-neighbor ({\em n.n.n.}) interaction, namely 
\begin{equation}\label{hamilt}
{\hat{\cal H}}=\sum_{i} \left \{ J_1{\hat{\bf S}}_i\cdot{\hat{\bf S}}_{i+1}
+J_2[1+(-1)^i\delta]{\hat {\bf S}}_i\cdot{\hat{\bf S}}_{i+2} \right \}~,
\end{equation}
where $J_1,J_2\geq 0$ are the nearest- and next-nearest neighbor
exchange couplings, respectively and $0 \leq \delta \leq 1$ 
is the bond alternation parameter. 
In the following, all the energies are expressed in units of $J_1$.
For $\delta=0$ the properties
of the model and the existence of a 
quantum phase transition at $(J_2/J_1)_c\simeq 0.241$ 
are well established.~\cite{affleck,j1j21d}
Below the critical point the system is in a Luttinger spin-fluid phase 
with a gapless spectrum, no broken symmetry, 
and a finite spin-wave velocity. Spin correlations are
characterized by a power-law decay defined by the same exponent of the
pure Heisenberg chain ($J_2=0$). 
In the regime of strong frustration, instead, the ground state is spontaneously dimerized,
two-fold degenerate and adiabatically connected to the Majumdar-Gosh exact
solution for $J_2/J_1=0.5$~\cite{mg,ss} with exponentially decaying spin correlations. 

The effect of the bond alternation on the {\em n.n.n.} interaction
has been recently investigated using field-theoretical 
approaches~\cite{voit,cinesini} with remarkable disagreement in the conclusions.
In fact, Chen and collaborators~\cite{voit} using bosonization and a 
renormalization group analysis claimed that in the limit of small frustration
the asymmetry in the {\em n.n.n.} integrals
destabilizes the isotropic Heisenberg fixed point leading to
a new phase with gapless excitations and vanishing spin-wave
velocity. On the other hand, with a similar bosonization approach Sarkar and 
Sen~\cite{cinesini}
found that bond alternation represents an irrelevant
perturbation in the regime of small frustration.
In this paper, using conformal field theory
we show that the {\em n.n.n.} spin-Peierls operator, i.e., the one associated
with the alternation in the {\em n.n.n.} exchange,
\begin{equation}\label{nnn}
{\hat O}_{n.n.n.}=\sum_{l} (-1)^l {\hat {\bf S}}_l\cdot{\hat{\bf S}}_{l+2},
\end{equation}
represents indeed an irrelevant perturbation for the Heisenberg
chain in the regime of weak frustration. Furthermore, using Lanczos
diagonalization technique we find that in the dimerized phase
the effect of the bond alternation is to reduce the spin gap 
leading therefore to a slight enhancement with $\delta$ of the critical value $(J_2/J_1)_c$ 
for the fluid-dimer transition. Finally, we present an accurate numerical
determination of the shift introduced by the bond alternation 
on the phase boundary between the Luttinger and the dimerized phase.
A preliminary account of this work was already presented in Ref.~\onlinecite{beto}.

In general, the relevance of the operator~(\ref{nnn}) for the 
Luttinger fixed point can be determined by calculating 
its scaling dimension $X$. This can be defined in terms of the size-dependence
of its generalized susceptibility
\begin{equation}
\chi_O = \frac{2}{L} \langle \psi_{0}| \hat{O} 
(E_{0}-{\cal\hat{H}})^{-1} \hat{O}|\psi_{0}\rangle\propto L^{2(1-X)}~,
\end{equation}
where $L$ is the number of sites of the ring, so that $X<1$ characterizes 
a relevant operator.~\cite{bibbia} Here $|\psi_{0}\rangle$ represents 
the unperturbed Luttinger ground state.
The calculation of $X$ can be done by identifying
the quantum numbers $\{j\}$ (referenced to the ground state with
energy $E_0$) of the intermediate states $| j \rangle$ appearing
in the Lehmann representation of $\chi_O$.~\cite{bibbia}
In fact, in a conformally invariant field theory, the scaling dimension $X$
of a given operator is related to the finite-size corrections of the 
energy of the lowest intermediate eigenstate $|j \rangle$ by 
\begin{equation}
\Delta E(L)=E_{j}(L)-E_0(L) = 2\pi v_s X/L~,
\end{equation}
where 
$v_s$ is the spin-wave velocity. In the case of the {\em n.n.n.}
operator~(\ref{nnn}), the quantum numbers of the intermediate states, 
referenced to 
the ground state, are the total spin ${\hat{\bf S}}=0$, momentum $k=\pi$ and 
even parity under spatial reflection ($l\to L-l$), $R=1$.
In all cases, the only difference with the nearest-neighbor ({\em n.n.}) 
spin-Peierls operator,
\begin{equation}\label{nn}
{\hat O}_{n.n.}=\sum_{l} (-1)^l {\hat {\bf S}}_l\cdot{\hat{\bf S}}_{l+1}~,
\end{equation}
the well-known relevant perturbation, is the spatial-reflection quantum number, 
$R=-1$.
The finite-size corrections can be computed either by Bethe {\em ansatz}
(for $J_2/J_1=0$) or by bosonization.
According to the standard notation, at the Luttinger fixed point, the 
effective Hamiltonian describing spin fluctuations
is given by a free boson theory:~\cite{affleckrev}
\begin{equation}
H_L = \frac {v_s}{ 2} 
\int dx \left \{ K\Pi^2(x) +\frac{1}{K} [\partial_x\Phi(x)]^2\right \}~, 
\label{lutto}
\end{equation}
where $v_s$ is the spin velocity, $\Phi(x)$ is the spin fluctuation field 
and $\Pi(x)$ is the conjugated moment. The dimensionless coupling constant $K$ 
is the key parameter of the bosonization theory, whose value can be analytically 
determined at $J_2=0$ by comparing the low-lying excitation spectrum of the 
Hamiltonian~(\ref{lutto}) with the Bethe {\em ansatz} solution of the lattice 
model: $K=(2-2\cos^{-1}\lambda/\pi)^{-1}$. 
Here $0\leq\lambda\leq 1$ is the coupling constant
defining the easy-plane anisotropy: at the spin-isotropic point $\lambda=1$,
leading to $K=1/2$.
The finite-size corrections of the lowest excitation energies of this 
Hamiltonian are given by:~\cite{bibbianoi}
\begin{equation}\label{eccito}
E_{n,m}(L)-E_0(L) = \frac {2\pi v_s} { L} \left ({n^2\over 4K} + Km^2\right )~.
\end{equation}
For each pair of integers $n$ and $m$ ($n=m=0$ being the ground state), a 
low-energy excited state with energy $E_{n,m}$ is defined by choosing 
appropriate boundary conditions for the fields:
\begin{eqnarray} \label{boundary}
\Phi(L)-\Phi(0) &=& n \sqrt{\pi}~, \\
\int\limits_{0}^{L} dx \Pi(x) &=& 2 m \sqrt{\pi}.
\end{eqnarray}
Using the relation between the boson fields and the physical electron operator
ones, it was then derived in Ref.~\onlinecite{affleckrev} that $n$ is the $z$ 
component of the spin, while $m+n$ is the momentum of the low-energy 
excitation in units of $\pi$. 
Following this work, it can be shown that the reflection symmetry ${\cal R}$ 
acts on the fields  as follows:
\begin{eqnarray} \label{reflection} 
{\cal R}^{\dag} \Phi(x) {\cal R} &=& -\Phi(L-x)~, \\
{\cal R}^{\dag} \Pi(x) {\cal R} &=& -\Pi(L-x)~, 
\end{eqnarray} 
implying that the quantum number $m$ changes sign under spatial reflection 
${\cal R}$, whereas $n$ does not, and, therefore, the ground state has $R=1$.
For $m \ne 0$, each pair $\pm m $ leads to two degenerate eigenstates 
with opposite reflections, $R =\pm 1$.

Clearly, the excitations of Eq.~(\ref{eccito}) do not exhaust the full energy 
spectrum of the Luttinger model because, on top of them, the usual boson
spectrum is also present:
\begin{equation}\label{bosons}
E_k(L) = v_s \sum_{k \ne 0 } |k| a^{\dag}_k a_k~,
\end{equation}
where $a^{\dag}_k$ is a boson field defined in terms of
$\Phi(x)$ and $\Pi(x)$, which satisfies
${\cal R}^{\dag} a^{\dag}_k {\cal R} =-a^{\dag}_{-k}$.

\begin{figure}
\vspace{-2.0cm}
\includegraphics[width=0.45\textwidth]{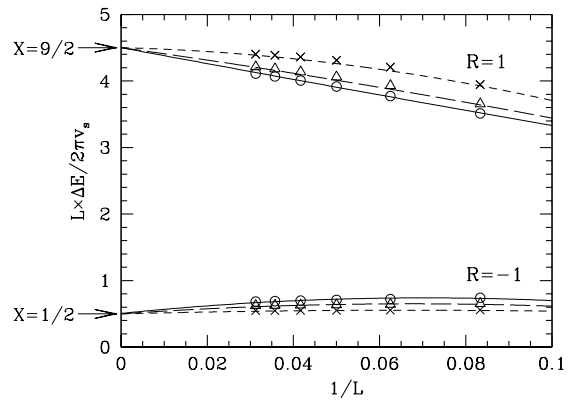}
\vspace{-0.4 cm}
\caption{\label{betofig}
Size scaling of the gap of the lowest eigenvalues of the isotropic spin-half 
$J_1{-}J_2$ chain with $S^z=0$, $k=\pi$, even parity under spin reflections,
and $R=\pm 1$ under lattice reflections: $J_2/J_1=0$ (circles), 
$0.1$ (triangles), and $0.2$ (crosses). Lines are guides for the eye.}
\end{figure}

For both the spin-Peierls operators previously defined, 
the appropriate choices of quantum numbers are $n=0$ (i.e., $S^z=0$)
and $m={\rm odd}$ (i.e., momentum $\pi$). 
The two lowest excitations with $m=\pm 1$ and $n=0$ can be combined 
to construct two states with defined reflection parity, $R= \pm 1$.
In the spin-isotropic case, $K=1/2$, the excitations with ($m=\pm 1$, $n=0$)
and ($m=0$, $n=\pm 1$) are degenerate, the latter ones corresponding to
a $R=1$ triplet (the spatial reflection does not change $n$). Therefore,
the state with $n=0$ and $R=1$ is the $S^z=0$ state of the triplet,
and the remaining state with $n=0$ and $R=-1$ has to be a singlet.
Analogously, all finite-$m$ excitations with $n=0$ are forbidden to be 
singlets with $R=1$ by the spin isotropy.  
In order to obtain a low-energy singlet excitation with $R=1$,
we have to combine the $m=1$ ($R=-1$) excitation with at least three 
$a^{\dag}_{k}$ bosons with zero total momentum. In fact one boson  
cannot have zero momentum and two bosons cannot change the
spatial reflection symmetry.
The minimum energy of such state is readily evaluated 
[with $k_1=k_2=2 \pi/L$, and $k_3=-(k_1+k_2)$] as 
\begin{equation} \label{lspectrum}
\Delta E(L)= 2\pi v_s(K+4)/L~,
\end{equation}
leading to the result $X=K+4$ for any operator with the same symmetries 
of the operator~(\ref{nnn}). 

Although the previous equation has been extracted in the isotropic case 
($\lambda=1$), where $K=1/2$, it is indeed valid in general for 
$0\le\lambda\le 1$, as can be verified, analytically, in the 
XY model (i.e., at $\lambda=0$) or, numerically, for $\lambda<1$ 
(not shown). 
The reason is that, in the entire gapless regime, the 
low-energy excitations are described by Eqs.~(\ref{eccito}) and~(\ref{bosons}).
Notice that, for $\lambda \ne 1$, the total spin is no longer a good quantum 
number, nevertheless the parity under spin-reflection
[$(S^x,S^y,S^z)\to (-S^x,S^y,-S^z)$] can still be used to label the 
low-lying states.

\begin{figure}
\vspace{-1.0 cm}
\includegraphics[width=0.45\textwidth]{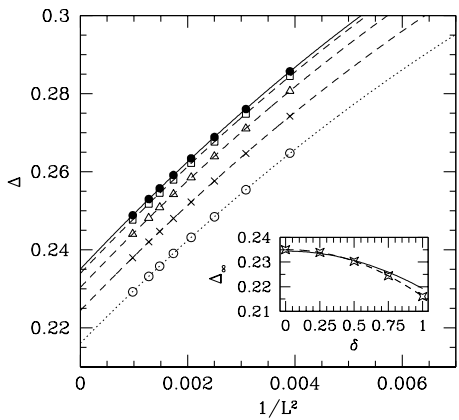}
\vspace{-1.2 cm}
\caption{\baselineskip .185in \label{gap}
Size scaling of the triplet gap at $J_2/J_1=0.5$ for $\delta=0$ (full circles),
$0.25$ (squares), $0.5$ (triangles), $0.75$ (crosses), and $1.0$ (empty circles).
Lines are quadratic fits. Inset: thermodynamic gap versus $\delta$ obtained 
by extrapolation (stars and dashed line) and with the variational approach 
of Refs.~\protect\onlinecite{voit,sen,ss} (continuous line).}
\end{figure}

The previous analysis shows that the finite-size spectrum of the Luttinger chain
does not contain low-lying singlets with $k=\pi$ and even parity under 
spatial-reflections, yielding to a critical exponent $X>1$ for the
operator~(\ref{nnn}), and therefore proving its irrelevance.
This is also illustrated in Fig.~\ref{betofig}, where we report exact 
diagonalization results in the weak frustration regime of the isotropic 
chain.~\cite{notesw} 
However, we stress that our conclusions follow directly from the bosonization 
technique and represent an exact result for the isotropic spin-half Heisenberg 
chain.

The irrelevance of the {\em n.n.n.} spin-Peierls perturbation in the regime
of weak frustration can be also confirmed numerically from the Lanczos 
calculation of the associated generalized susceptibility. 
As explained in Ref.~\onlinecite{plaquetto}, this can be done by adding to the 
Hamiltonian the considered perturbation $\eta \hat{O}$
and then computing the ground-state energy per site $e(\eta)$ 
for few values of $\eta$ in order to
estimate numerically $\chi =-d^2e(\eta)/d\eta^2|_{\eta=0}$ as the limit
$ \chi = \lim_{\eta \to 0} {2(e_0-e(\eta))}/{\eta^2}$.
For instance, for the $28$-site chain at $J_2/J_1=0.1$, this procedure gives
$\chi_{nn} \simeq 6.4$ and $\chi_{nnn} \simeq 10^{-6}$, for the {\em n.n.} 
and {\em n.n.n} spin-Peierls operator, respectively.
This fact indicates that the response of the system to a {\em n.n.n.} 
dimerization is more than six orders of magnitude smaller than the response
to the standard {\em n.n.} dimerization.

\begin{figure}
\includegraphics[width=0.45\textwidth]{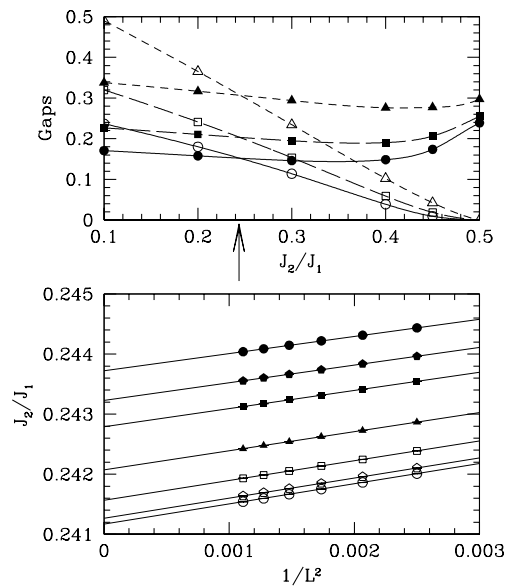}
\vspace{-0.1 cm}
\caption{\baselineskip .185in \label{alphacrit}
Upper panel: singlet (open dots) and triplet (full dots) gaps versus
$J_2/J_1$ for $\delta=1.0$, $L=12$ (triangles), $18$ (squares), 
and $24$ (circles). The arrow indicates the extrapolated value
of $J_2/J_1$ at the level crossing. Lines are guides for the eye.
Lower panel: size scaling of $J_2/J_1$ at the level crossing for (from below)
$\delta=0, 0.2, 0.4, 0.6, 0.8, 0.9, 1.0$. Lines are quadratic fits.}
\end{figure}

Entering the regime of strong frustration, the effect of the {\em n.n.n.}
alternation is to reduce the spin gap $\Delta$ as it is shown in Fig.~\ref{gap}
at $J_2/J_1=0.5$.
As it has been already pointed out,~\cite{voit,cinesini}
at the Majumdar-Gosh point ($J_2/J_1=0.5$)
the presence of the {\em n.n.n.} spin-Peierls dimerization does not alter the 
ground-state manifold which is spanned by the two well-known dimerized 
states, with energy $E_0=-3/8 N$, independent of $\delta$.
Nevertheless, the excited states are not known 
exactly so that we have calculated the spin gap using the Lanczos technique.
By extrapolating the finite-size data according
to the expected law,~\cite{scaloppo} $\Delta(L)= \Delta_\infty + a/L^2+b/L^4$,
we have obtained the $\delta$ dependence of the thermodynamic gap, which is 
shown in the same figure.
Our results are in good agreement with the
variational calculation of Refs.~\onlinecite{voit,sen,ss}, and indicate a 
$8\%$ spin-gap reduction from the Majumdar-Gosh model for $\delta=0$ 
($\Delta= 0.235$) to the so-called saw-tooth chain for $\delta=1$ 
($\Delta=0.216$).~\cite{note2} 

\begin{figure}
\vspace{-2.0 cm}
\includegraphics[width=0.45\textwidth]{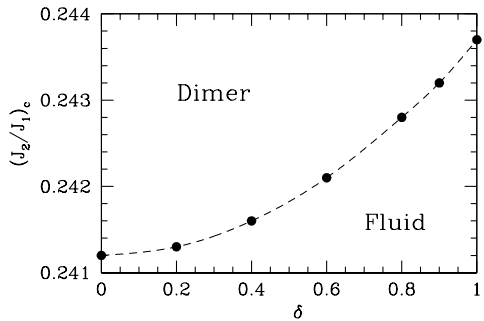}
\vspace{-1.6 cm}
\caption{\baselineskip .185in \label{asl3} \label{phasediag}
Critical $J_2/J_1$ ratio for the fluid-dimer transition versus $\delta$.}
\end{figure}

The reduction of the spin gap in the dimerized phase 
induced by the {\em n.n.n.} 
bond alternation leads to an increase of the critical value $(J_2/J_1)_c$ for
the fluid-dimer transition. 
As originally suggested by Castilla and collaborators,~\cite{emery1d}
the phase boundary between the two phases can
be calculated very accurately by exploiting the change in the 
excitation spectrum occurring at the critical point.
In fact, below the critical point the system displays
quasi-long-range antiferromagnetic order: in this case,
gapless excitations are obtained by creating low-energy
spinons both in triplet (spatially even) and singlet (spatially odd) states,
the former having the lowest energy on finite size.
In the dimerized phase, instead, the triplet excitations acquire a gap
so that the ground state remains two-fold degenerate in the singlet sector,
leading to the dimerized ground state which breaks the translational invariance.
As a result, the critical coupling $(J_2/J_1)_c$ can be determined
by performing a size scaling of the value $J_2/J_1$ 
where the level crossing between the low-lying singlet and triplet 
occurs (see Fig.~\ref{alphacrit}). 
With this technique, using exact diagonalizations up to $30$ sites,
it is easy to obtain very accurate determinations of the critical 
value $(J_2/J_1)_c$. 
For instance, in absence of {\em n.n.n.} bond alternation,
we obtain a critical value $(J_2/J_1)_c$ in agreement with previously given 
estimates.~\cite{j1j21d,emery1d}
The resulting phase boundary for the fluid-dimer transition is shown
in Fig.~\ref{phasediag}, indicating that for $\delta=1$ the 
critical value of the frustration ratio for the
fluid-dimer transition is only increased by about $1\%$.

In conclusion, we have studied the effects of a next-nearest-neighbor 
spin-Peierls dimerization on the ground state of the spin-half
$J_1{-}J_2$ Heisenberg chain, using conformal field theory and 
exact diagonalization. We have found that, despite the conclusions of a recent
study based on bosonization and renormalization group approaches,~\cite{voit} 
the bond alternation represents an irrelevant perturbation in both the 
regime of strong and weak frustration.
In particular, as predicted by Sarkar and Sen,~\cite{cinesini} 
the perturbation does not change the nature of the excitation spectrum,
leading only to a weak reduction of the triplet gap in the dimerized phase, and,
therefore, to a slight enhancement of the critical value $(J_2/J_1)_c$ 
associated to the fluid-dimer transition. 

We would like to thank D. Sen for having attracted our attention
to this model and for useful correspondence.  
L.C. and A.P. acknowledge kind hospitality at SISSA.
This work was partially supported by MIUR (COFIN01) and INFM (PAIS-MALODI).
L.C. was supported by NSF grant DMR-9817242.



\begin{thebibliography}{99}
\bibitem{notesw} The spin-wave velocity $v_s$ is known exactly 
only for $J_2=0$, where $v_s=\pi/2$. However, in the Luttinger regime 
it can be obtained numerically from the finite-size corrections to the 
ground-state energy according to the relation, $E_0(L)=e_0 L-v_s\pi/6L$, 
where $e_0$ is the thermodynamic ground-state energy per site. See for
instance, K. Nomura and K. Okamoto, J. Phys. A {\bf 27}, 5773 (1994).
\bibitem{note2} The value for the saw-tooth chain ($\delta=1$) is in agreement 
with the recent calculation of Ref.~\onlinecite{saw}.
\end{thebibliography}
\end{document}